\documentstyle[twocolumn,prc,aps,epsf,epsfig]{revtex}
 \begin{document}
 
\newcommand{\be}{\begin{eqnarray}}
\newcommand{\ee}{\end{eqnarray}}
\twocolumn[\hsize\textwidth\columnwidth\hsize\csname@twocolumnfalse\endcsname
%\preprint{SUNY-NTG-xxx}
\title{
Matter-induced modification  of resonances at RHIC freezeout
}

\author{  E.V. Shuryak and G.E.Brown  }

\address{
Department of Physics and Astronomy, State University of New York, 
Stony Brook NY 11794-3800, USA
}

\date{\today}
\maketitle

\begin{abstract}
We discuss the physical effects causing a
modification of resonance masses, widths  and even
shapes in a dilute hadronic gas
at late stages of heavy ion collisions.
We quantify the conditions at which  resonances are produced at RHIC,
and found that it happens at $T\approx 120 \, MeV$.
Although in the pp case the
 ``kinematic'' effects like thermal weighting of the states
is sufficient, in AA we see a clear effect of 
dynamical interaction with matter, both due to
a variety of s-channel resonances
and due to  t-channel scalar exchanges.
The particular
quantity we focus mostly on is the
$\rho$
meson mass, for which these dynamical effects lead to about -50 MeV
shift, on top of about -20 MeV of a thermal effect: both
agree well with preliminary data from  STAR experiment at RHIC. 
We also predict a complete change of shape
of $f_0(600)$ resonance, even by thermal effects alone.

\end{abstract}
\vspace{0.1in}
]
\begin{narrowtext}
\newpage
%\vskip 2pc]

\section{Introduction}
  The first experimental data from the Relativistic Heavy Ion Collider (RHIC)
at Brookhaven National Laboratory have opened a new page in studies
of hot/dense hadronic matter. The magnitude of collective effects
strongly suggests rapid equilibration in the system. Strong
jet quenching, its angular dependence and the absence of
compensating jet all suggest the surface 
emission of jets. Why the bulk of the matter is effectively 
black, even for $\sim 10
\, GeV$ hadrons/jets, remains
 unknown at this time. Another unsolved RHIC puzzle is 
the apparent large fraction of baryons at large $p_t$.
 All those  phenomena at large
transverse momenta are presumably 
related to properties of extremely hot/dense
hadronic matter, preceding formation of the Quark-Gluon Plasma.

  In this paper we would however approach the phenomena
from the opposite side, focusing instead 
on the most dilute stages of the collision close to the  freezeout.
It has been argued
over the years that
under such conditions hadrons and especially
short-lived resonances should be modified, with
 shifted
mass, increased width and even significantly changed shape. 
The most dilute stage
of the collisions, known as {\em kinetic} freezeout, is much more
dilute
than nuclear matter. In addition,
 most of
 the particles are Goldstone bosons, $\pi,K,\eta$, which interact
 weakly
at low energies.
At such low density matter the proposed modifications are expected
to be small, although
 quite observable. They should naturally be studied to set a
 benchmark,
before extrapolation to denser matter close to the QCD phase transition,
at which more species of hadrons are actually produced (the chemical
freezeout).

Recently  the STAR experiment at RHIC reported 
observed signals of a variety of mesonic resonances 
 such as $\rho,\omega,K^*,f_0(980)$ \cite{STAR_Fachini}
 and $ \phi$ \cite{STAR_Yamamoto} as well as baryonic
resonances (which we will not discuss in this work). 
A  data sample reported
corresponds to  mid-central (40 to 80 percent of the hadronic cross section)
AuAu collisions at $\sqrt{s}=200 \, GeV$ is shown in Fig.\ref{fig_STAR_res}.
(They are not yet acceptance corrected, so the numbers extracted from
them should not be directly compared to our calculations below.)
 One can see from comparison of pp and AuAu spectra
that the shape of the $\rho-\omega$
pair of resonances in the $\pi^+\pi^-$ channel is quite different from
its classic shape in $e^+e^-\rightarrow \pi^+\pi^-  $ or $\tau$
decays.
Unlike those leptoproduction reactions, in which
the  masses of $\rho$ and $\omega$ are nearly
degenerate, in pp and even more so in 
 AuAu collisions  the $\rho$ mass is
shifted  downward, while the $\omega$ mass stays the same.
The (preliminary) fits \cite{STAR_Fachini} gave 
$m_\rho=0.698 \pm 0.013 \, GeV$ in AuAu and $0.729 \pm 0.006 \, GeV$
in pp, with about unchanged width.

\begin{figure}[h!]
\centering
\includegraphics[width=8cm]{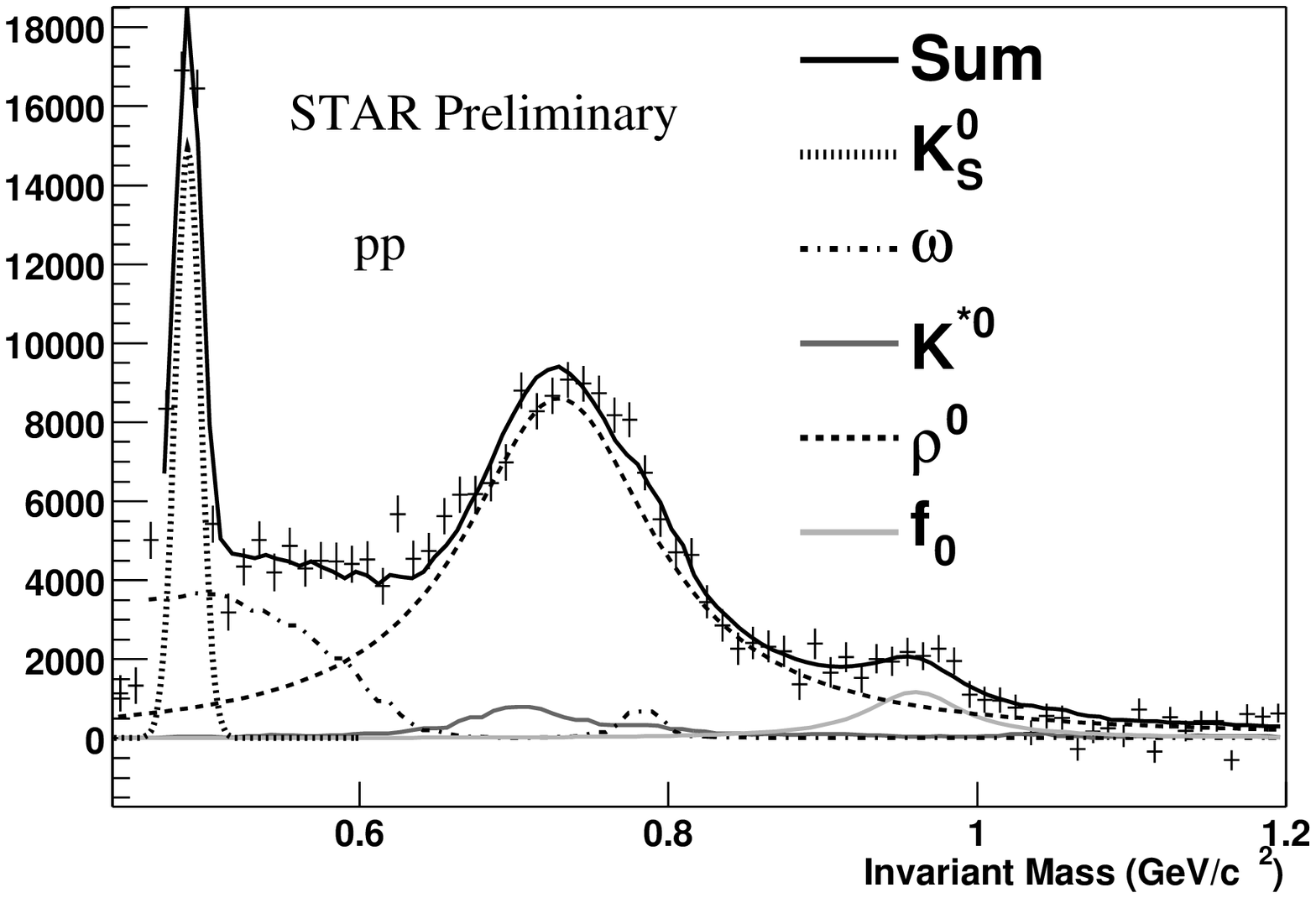}
\includegraphics[width=8cm]{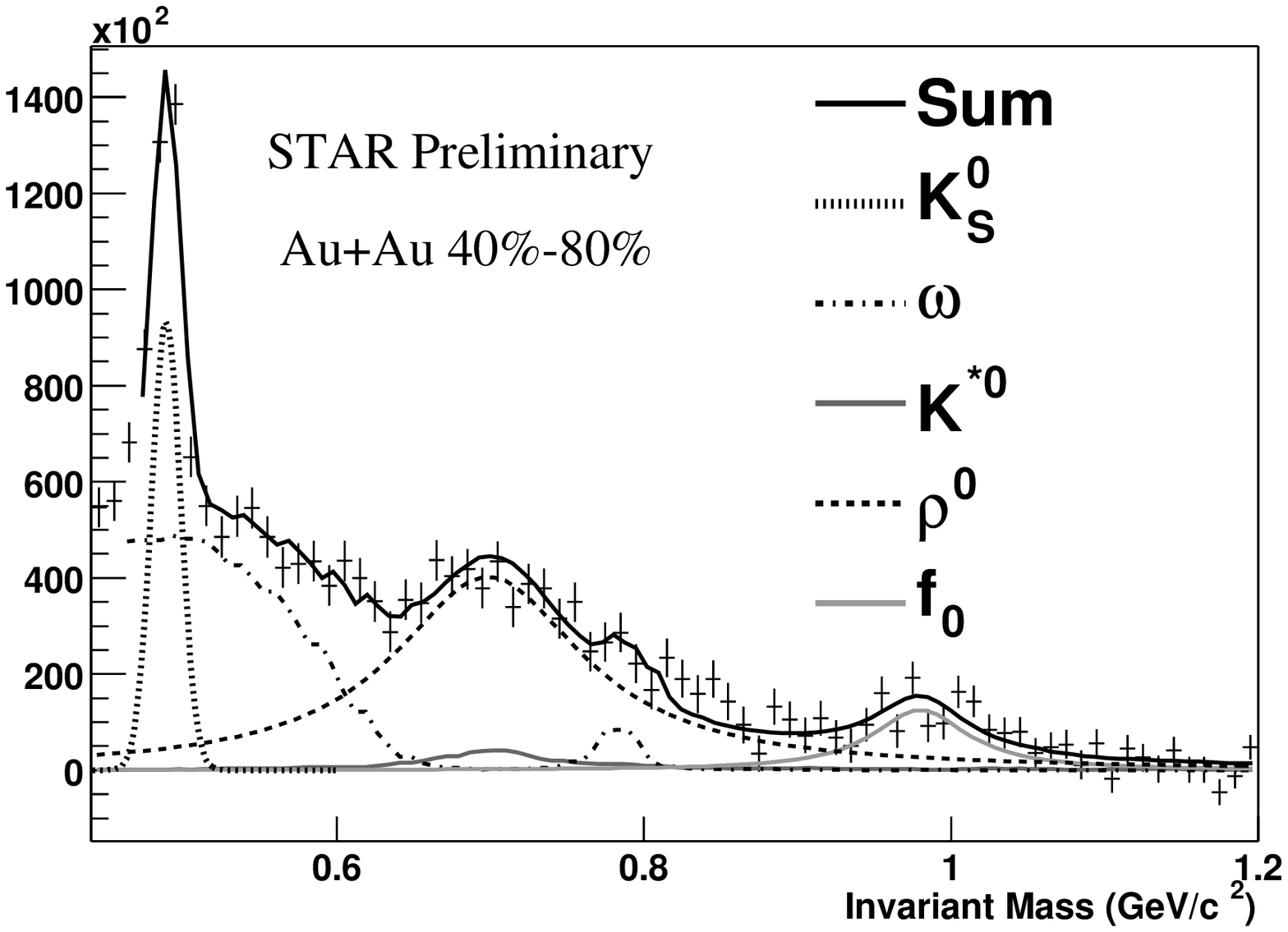}
\caption{\label{fig_STAR_res} 
The invariant mass distributions in pp and mid-central AuAu
 of the $\pi^+\pi^-$ system, with a
transverse momentum cut
$0.2 < p_t < 0.9 \, GeV$. The lines indicate contribution of particular
resonances according to some model we do not discuss in this work.
}
\end{figure}

\subsection{esonance mass modification: theory}

In general, 
resonance modification
in hadronic matter results from  mutual
interaction, and is analogous to the well known phenomena in  atomic
physics known as spectral line shifts and  broadening.

In the first order in density of some a hadronic species
$n_i$, a modification 
of a hadron of kind $j$ can be expressed in terms of their {\em forward
scattering amplitude} $M_{ij}(t=0,s)$.
It is similar to many known phenomena
 such as e.g. a modification
of the photon dispersion law when it propagates though 
 glass.
Note that the scattering amplitude is complex, and that this approach
gives
both the real and imaginary part of the dispersion law modification, also known as
the optical potential. Note also that if the dependence on momenta
of both particles is strong, appropriate integration with the thermal
weights
should be performed. 

 Unfortunately, only partial information about the
scattering amplitude can be obtained from the experiment.
There are two
major theoretical approaches to the issue discussed in literature,
 to be called an $s-channel$ and a $t-channel$ one.

The former approach assumes that the scattering amplitude is dominated by
s-channel resonances which are known to decay into the $i+j$  channel.
For most mesons such as
$\pi,\omega,\rho,K$  in a gas made of pions such calculation has been
made \cite{Shu_pots,Shu_Tho,Kap_etal,RG} 
It is usually true that
the single most important part of scattering
amplitude is the lowest resonance. A general rationale for that
is that higher resonances typically decay into states with higher
multiplicity, with the 2-body channel in question having a small branching ratio. 
In the case we will discuss
most
below -- the $\rho\pi$ scattering -- it is the axial resonance  $a_1$.
 Generic diagrams for such processes look like
that depicted
in Fig.\ref{fig_diagram}(a): 
 and in dilute matter the modification of the $\rho$
is simply proportional to the appropriately weighted pion density (see below).

 Such an approach is of course rather limited, applicable only for
dilute matter. Indeed, if one would like to follow it to second order
in density  with a double collision Fig.\ref{fig_diagram}(c), one
should
be able to extrapolate the scattering data off-shell (an intermediate
line with an asterisk). 
The same problem appears if one may think about loop corrections.
Furthermore, generic 3-body
collisions cannot be  accessed experimentally,
only their
part  due to intermediate resonances with the known 3-body decays
(see Fig.\ref{fig_diagram}(d)) can.

 The predicted \cite{Shu_pots} $\rho$ mass shift due to the $a_1$ in
 an equilibrated\footnote{That is, with zero pion chemical potential. The
 calculations
below correspond to significant $\mu_\pi$ which of course increase the
 effect.}
 pion gas
was found to be small, not exceeding
 -10 MeV.  The
sign of this effect can be explained as follows:
 states of the same quantum numbers
repel each other if mixed, and thermally populated
$\rho\pi$ states have total energy mostly below the mass of $a_1 $.

\begin{figure}[h!]
\centering
\includegraphics[width=8cm]{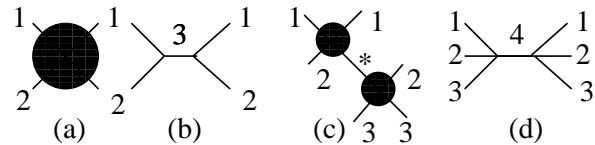}
\caption{\label{fig_diagram} 
Forward scattering amplitude for particles of type 1 and 2 (a) can be
approximated by intermediate resonance 3 (b). Even the double 2-body 
scatterings (c) include an off-shell particle (asterisk) which cannot be
calculated without assumptions. An exceptional case is a single
resonance 4 which has an observable  3-body  decay (d).
}
\end{figure}

Similar effects due to scattering on nucleons have been studied e.g.
by Rapp, Wambach and collaborators, see \cite{CRW} and references
therein. There is however an important difference in sign of the effect.
 The major resonance relevant for
 $\rho N$ scattering on nucleons, $N^*(1520)$, is not only below the
thermally populated states but even below 
$m_\rho+m_N$: thus it pushes the $\rho$ mass upward.
The lower excitation, which is the  $N^*(1520) N^{-1}$ state,  is at
the same time
pushed  down. Another important effect discussed a lot in these works is
significant broadening of both states, leading to a complicated quasi-continuum
spectrum in place of two well separated states.

 Let us now turn to
the second $t-channel$ approach, attributing
the mass shifts to mutual attractive interaction between hadrons,
such as t-channel exchange of a scalar isoscalar $\sigma$ meson,
see Fig.\ref{fig_exchange}.
We first comment on a significant controversy related to its association
with the issue of depletion of the chiral condensate and chiral
restoration
phenomenon. Let us try to explain briefly what is its status at this point.

\begin{figure}[h!]
\centering
\includegraphics[width=2cm]{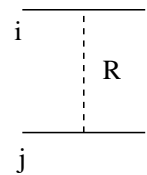}
\caption{\label{fig_exchange} 
Forward scattering amplitude for particles of type i and j exchanging
the ``radial'' scalar $R$.
}
\end{figure}

On the one hand, a modification of the quark condensate can 
be calculated at small temperatures 
 from general chiral theory  \cite{Leutwyler} without a problem, the
 result is
\be \label{eq_cond}
<\bar q q>(T)=<\bar q q>(0)(1-{T^2\over 8f_\pi^2}- {T^4\over
  384f_\pi^4}+...)\ee
 The hadronic masses appear due to chiral symmetry breaking,
and in  models like NJL it happens
via the so called constituent quark mass which  is
directly proportional
to the quark condensate.  Therefore naively 
\be {m_\rho^* \over m_\rho} \sim { <\bar q q>^* \over <\bar q q>}\ee
which is
sometimes referred to in the literature  as a ``Nambu scaling''.

Here and below in-matter quantities are marked by the asterisk.
 Brown and Rho  have  proposed 
another scaling ( see the
latest work \cite{BR_scaling} for earlier references and explanations) 
\be \label{eq_BR}
{m_\rho^* \over m_\rho} \sim {f_\pi^* \over f_\pi}\sim ({ <\bar q q>^*
  \over <\bar q q>})^{\alpha}\ee
with  $\alpha=1/2$ for low densities,

The problem however is that 
such relations  are not 
respected by low-T  chiral calculations. In particular,
the   
$O(T^2/f_\pi^2)$ effect which is present
in the quark condensate (\ref{eq_cond}) as well as in 
of the vector and axial correlation functions \cite{DEI},
 is absent in physical effects like
the total energy of the pion gas \cite{Shu_80} or in the mass
shifts, which appears only in order $O(T^4)$, see e.g. \cite{EI}.
More generally, it is clear that a t-channel exchange  of a sigma meson 
between a pion (with a $non$-derivative coupling)
and any other hadron would violate the Goldstone theorem and
lead to a non-zero  pion mass. This contribution must therefore cancel
out with some other diagrams, as seen in explicit calculations in many
models.

  Predictions for 
next order   $O(T^4)$ mass shifts are model-dependent and there
is no theoretical consensus even about the $sign$ of the contribution.
The $O(T^4)$  calculation
performed
in \cite{EI} has  predicted that both $\rho,a_1$ masses
 decrease slightly,
 in equilibrium $T=150 \, MeV$ was found to be
$\delta m_\rho^{EI}=-6.4 MeV$. 

In contrast to that, Pisarski \cite{Pisarski} has
built a set of simple effective models incorporating $\pi,\rho,a_1$
interactions (which all of course respect chiral symmetry) 
and concluded that no prediction can be really made. 
In one of them, enforcing VDM at any T, the result is
that the $\rho$ mass should $grow$ with T significantly, joining 
 $a_1$ at at chiral
restoration point at the following intermediate mass 
\be m^2_\rho(T_\xi)= m^2_{a_1}(T_\xi)={2\over 3}m^2_\rho(0)+{1\over
  3}m^2_{a_1}(0)=(962 \, MeV)^2 \ee

  In contrast to scalar exchanges with pions, such
 exchanges between other hadrons definitely exist.
(For nucleons those 
are in fact responsible for a large portion
of nuclear forces and  binding.)
The effect of nucleon density on quark condensate  related with the
nucleon sigma term \cite{DL} is large even at nuclear matter density
\be { <\bar q q>^* \over <\bar q q>}=1-{\Sigma_{\pi N} n \over f_\pi^2
  m_\pi^2}+... \approx 1-.36 {n\over n_0}\ee
where $n_0$ is the nuclear matter density.
The resulting BR scaling prediction 
\be m_\rho^*=m\rho (1- 0.36 {n\over n_0}+...)^{1/2}\ee
leads to good agreement with experimental information
on scattering on nuclei.

This apparent contradiction is discussed in multiple papers. We would follow
\cite{CEG} where
 it has been especially clearly emphasized that one may think of
$two$ different scalar fields which can be exchanged. One is the old
$\sigma$
of the linear sigma-model, a chiral partner of the pion. The other
we will call R \footnote{From 
the radius - in \cite{CEG} it was called $\theta$ 
which may be mistakenly taken as an angle or
may be confused with the widely used
 $\theta$ field, the U(1) pseudoscalar. $R$ is the $\chi$ of Brown and
Rho \cite{BR}.} 
\be R^2=\sigma^2+\pi^2 \ee
The main difference between VEVs $<\sigma>$ and $<R>$ is that the former
gets the pion loop $O(T^2)$ correction while the latter does not
since in the excitations along the chiral circle the radius does not change.
In a different language, 
the interaction Lagrangian describing
how $R$ interacts with pions is different from a sigma model one, it 
 has a derivative coupling like for any other particle.
Therefore  R-exchanges with pions
does not give them a mass and does not
 violate the Goldstone theorem; there is no need for cancellations.
The mass shifts for N and $\rho$
can then in the first order be written as R-exchange diagram or
\cite{CEG}
\be \label{eq_Rmassshift}
\delta M_N={3\over 2} \delta M_\rho=-{g_0^2 n \over m_R^2} \ee
In this simple picture $g_0=M_n/f_\pi\approx 10$ and R can be viewed
as the Walecka scalar meson, see also \cite{BR} where $R$ is
denoted as $\chi$.

\subsection{Resonance mass modification: experiments}
We will not go into long history
of detailed $\rho$-meson shape, but simply state that the difference between
its appearance in elementary
reactions ($e^+e^-\rightarrow \pi^+\pi^-  $ or $\tau\rightarrow
\nu_\tau \pi^+\pi^- $ decays) and  hadroproduction reactions is by now well
known.
The latest Review of Particle Properties (RPP) \cite{RPP}
averages the rho mass for these two classes of experiments separately,
with a clear systematic difference of the order of 10 MeV between
them:
\be m_\rho^{leptoproduced}=775.9\pm .5 \, MeV,\ee
\be \label{eq_hadroproduced}
m_\rho^{hadroproduced}=766.5\pm 1.1 \,MeV \ee.
We return to discussion of this effect in section \ref{sec_Boltz} below.

A   $\rho$ mass shift $\delta m_\rho=160 \pm 35 \, MeV$ observed in nuclear experiments has been
reported by Lolos et al \cite{Lolos} in the $^3 He (\gamma, \rho^0)
ppn$
reaction. An even larger shift has been reported by Huber et al
\cite{Huber} when
even lower $\gamma$ energies were used.
Although $^3 He$ is not a dense nucleus, with average
density
of about half of nuclear matter, the kinematic of this experiment is 
sub-threshold for one nucleon. It was therefore argued %\cite{MEricson} 
that the 
emerged
 $\rho$ is actually in the vicinity of at least two nucleons, so that
the local
density is large. 

  Heavy ion dilepton experiments at SPS, such as HELIOS3,NA38 and
  especially
CERES have indeed found that the vector spectral density of hot/dense
  matter
is very different from that observed in elementary pp collisions.
While in the latter case there exist a clear gap between
  $\rho,\omega$ mesons and low-mass pairs originated from Dalitz
  decays
of $\eta$ and other light mesons, 
no such gap is present  in heavy ion collisions and a
 continuum of excited  states is seen down to the invariant mass
of $\sim 400 \, MeV$.  Although these results agree with
various calculations such as the Rapp-Wambach ones mentioned above,
these dilepton data are  still very limited
statistically and have very crude mass resolution which does not make
  it possible
to see the $\rho$ modification or even separate
it from  the $\omega$ or even
 the $N^*(1520) N^{-1}$.    

The situation is very different for STAR experiment we mentioned in
the Introduction, in which the shifts are smaller and the width apparently
is not growing, while
the invariant mass resolution is
excellent and all resonances are seen as clearly resolved separate peaks. 

\section{Modification of $\rho$ meson at RHIC}
\subsection{Kinetic freezeout at RHIC}
  Early papers on statistical and hydrodynamical models
of multiparticle production had assumed complete equilibrium 
matter untill the final freezeout.
Clear separation between two  different stages, the
 chemical (at $T_{ch}$) and kinetic (at $ T_k$) 
freezeouts has been made in the mid-1990's. Basically it reflects
very different rates of elastic and inelastic reactions at low 
energies dominating rescattering of secondaries.

In particular, in ref.  \cite{HS}   the
following 
argument has been put forward: 
at AGS/SPS energies most of the observed
 velocity of collective radial flow can
$only$ develop in between the two, at  $T_{ch} < T< T_k$.

In that paper also the following natural definition of the kinetic freezeout
condition has been given: a secondary (say a pion) emitted at this
time has equal chances to be either rescattered or escape:
\be \label{freezeout_cond}
P_{escape}=P_{rescattering}={1\over 2} \ee
which we will be using below.

At the particular conditions we are interested in, at RHIC $\sqrt{s}=200
\, GeV$, the numerical
values of the parameters are as follows. The rapidity density $dN/dy$
at mid-rapidity of central AuAu collisions
for relevant
particles are about 300 for each species of pions, 29 protons and 22 
antiprotons. These numbers have large feed-down from resonance decays,
which should be accounted for when their
 ratios are used to derive the
chemical freezeout conditions. From 
\cite{STAR_Fachini} it was determined that those are about
\be \label{eq_chem}
T_{ch}=170\, MeV\,\,\,\, \mu_b=28\, MeV\ee
By definition of chemical equilibrium, all other chemical
potentials for non-conserved charges are zero at this stage. 

After inelastic reactions are frozen and only elastic re-scatterings
take place, the number of particles of each species are separately
conserved.
Therefore chemical potentials for all species $\mu_i$ become
$non-zero$ at the hadronic stage, 
see  details  in refs\cite{Bebie,HS,Teaney,Rapp}. 
The only important numbers for   our discussion below are  the
value of the kinetic freezeout temperature and the values of chemical
potentials for pions and nucleons. Assuming entropy conservation
(adiabatic expansion) and natural conservation of baryons
one can determine all parameters along a particular cooling path, provided
one point on it is known. Using   (\ref{eq_chem}) one gets it, and
using either kinetics, or hydro-based fit to data, one finds that
the kinetic freezeout at RHIC happens at 
\be \label{cond_kinetic}
T_k\approx 100 \, MeV, \,\,\, \mu_\pi\approx 81 \, MeV \nonumber \\
\mu_N\approx 380 \, MeV \,\, \mu_K\approx 167 \, MeV\ee
This translates into the following densities of the pions (all 3 of
them)
\be n_\pi\approx 0.06\, fm^{-3}  \ee 
With $\bar p/p\approx 0.75$ we get 
the following density of all nucleons plus anti-nucleons together
\be n_{N+\bar N}\approx 0.0075 \, fm^{-3} \ee
or only 1/20 of the nuclear matter density. Accounting for all
baryonic
 resonances will increase them by nearly a factor 3 sand bring them
in agreement with measured final yields mentioned above. 

The next issue we discuss is how the freezeout  conditions
for most particles (pions) relate to those for {\em resonances}.
When  those are actually observed as peaks in the invariant mass
distribution of several secondaries, it is only possible
when none of the secondaries have been rescattered. Suppose for
simplicity the resonance is produced and decays into the
same mode with N identical 
secondaries of type i (N=2 for $\rho\rightarrow 2\pi$). Its observable
production
at time t is proportional to the following combination
of the density needed for production and the probability that all
secondaries do escape, namely 
\be P(t)\sim \left[n_i(t)exp(-\int_{t+\tau}^\infty dt'\sum_j
n_j(t')
<\sigma_{ij} v_{ij}>)  \right]^N\ee 
where $n_i(t)$ is time-dependent density of 
relevant secondary in matter, and $<\sigma_{ij} v_{ij}>$
are thermally averaged cross section and velocity for scattering on
type-j
particle and $\tau$ is the resonance lifetime.

This expression can be simplified significantly, provided
the lifetime is ignored, and the matter expansion
is parametrized by a single power behavior $n_i(t)\sim 1/t^a$.
The cross section is approximately time-independent and can be eliminated
in favor 
 of the freezeout time, using its definition (\ref{freezeout_cond}),
leading to 
\be \label{eq_time_distr}
P_{resonance}(t)\sim \left[t^{-a}exp(-ln(2)({t_k\over t})^{a-1}
\right]^N \ee 
which for any N has the maximum at
\be  t_{res}=t_k \left[ln(2)*(1-1/a) \right]^{1\over a-1} \ee
As hydro simulations show\footnote{We thank P.Kolb who provided
plots from which the index value was estimated.} the value of the index 
is actually not constant over time; it changes from the Bjorken a=1
at early time of 1d expansion to a value $a\approx 2.5$ at relevant times $t\sim 10
\, fm/c$, not quite reaching ultimate 3 for 3d  expansion.
For this index,  
$t_{res}\approx 0.6 t_k$.  The  distribution (\ref{eq_time_distr}) for N=2 is plotted
in Fig,\ref{fig_res_time}, and one can see that the peak is rather
sharp. 

\begin{figure}[h!]
\centering
\includegraphics[width=6cm]{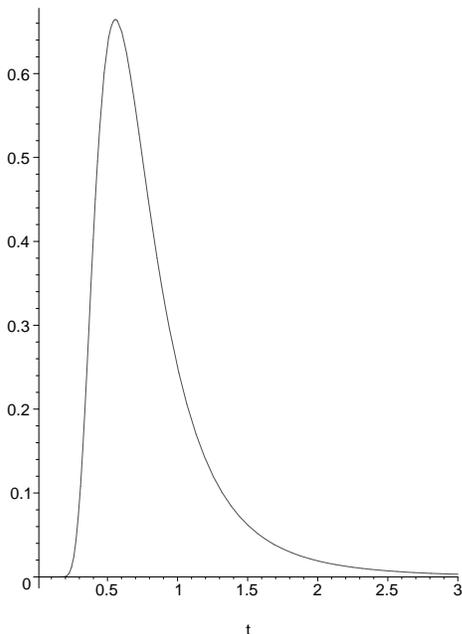}
\caption{\label{fig_res_time} 
Probability distribution over the 2-pion resonance production time t, 
in units of pion freezeout
time $t_k$.
}
\end{figure}

Finally, one can deduce from this time the relevant temperature
 using a  power-fit
for density at the hadronic stage $n\sim T^5$ which gives
\be {T_{res}\over T_k}=\left({t_{res}\over t_k} \right)^{-a/5}\approx 1.2\ee  
In effect, we found that the observed resonances are produced
 somewhat
earlier and a bit higher
temperature $T_{res}$ compared to the kinetic production time, for all N.  
The conditions on the same adiabatic path at such temperature are:
\be  \label{cond_res}
T_{res}\approx 120 \, MeV, \,\,\, \mu_\pi\approx 62 \, MeV \nonumber \\
\mu_N\approx 270 \, MeV \,\, \mu_K\approx 115 \, MeV\ee
with densities about 1.4 higher than at the kinetic freezeout.

\subsection{Elementary effects due to the heat bath}
\label{sec_Boltz} 
Processes like $\pi\pi$ scattering 
  in an equilibrium heat bath can be
represented as follows
\be 
{dW}= \left[ f(p_1)f(p_1) \right]|M|^2d\Gamma \left[(1\pm f(p_3))(1\pm
  f(p_4))
\right]
\ee
where (i) the first two factors are thermal occupation factors for the
  incoming particles; (ii) $|M|$ is the matrix element of the
  re-scattering,
 (iii) is the phase space and finally
(iv)  represent final state corrections for induced radiation, $\pm$ for
final  bosons/fermions; (iii)  In dilute matter one can presumably 
use unmodified matrix element and the phase space.

The socalled ``elementary'' hadronic reactions like pp collisions at
high energies are in fact very complex phenomena, which are far from
being understood. Unlike heavy ion collisions, those do not show
collective phenomena like flow and one may think they certainly 
do not contain equilibrated hadronic matter. And yet, 
as noticed back in 1960's by Hagedorn and others, the particle
composition in pp and even $e+e-$ collisions can be well reproduced by
statistical models, see e.g. recent comparison \cite{Beccatini}.
The fitted temperature is about the same as the chemical freezeout
temperature $T_{ch}$ mentioned above.
 Two- and even many-body
distributions
 follow thermal distributions very well, as data showed already in 1970's.
So it is natural\footnote{The authors thank P.Braun-Munzinger who
suggested this idea to them.
} to apply it to 2 pions which are forming the $\rho$ meson
in question as well. 

 Now, if two initial pions come from the heat bath even in pp,
 the initial thermal factors should be present. At the time the
resonance decays, on the other hand,
 the secondaries making that heat bath are already
sufficiently far away. From this
it follows that (i) the resonance decay happens in vacuum, its matrix element
is unmodified, 
and (ii) the two final state induced radiation factors are absent.

The  Boltzmann factor is of course a function of the energy
in the matter rest frame, not in the resonance rest frame. However
as a simple first approximation one may neglect the thermal motion of
rather heavy resonances and assume them to be at rest in the heat bath.
This simply combines the Boltzmann factor $exp(-M/T)$ containing
the invariant mass $M^2=(p_++p_-)^2$ with the
(appropriately modified for p-waves if needed)  Breit-Wigner matrix element.

The position of the maximum can then be obtained in a simple analytic way,
provided the resonance width is large enough. Approximating the resonance by
\be {exp(-M/T)\over 1+(4/\Gamma^2)(M-m_{res})^2}\approx \ee
$$ exp[-({4\over \Gamma^2})(M^2-2M m_{res}-M^2+{2M\Gamma^2\over 8T}]
$$ 
one gets the
mass shift due to (sufficiently large)
temperature T to be approximately
\be \delta m_{max}\approx -{\Gamma^2 \over 8 T}\ee

For $\rho$ emitted at $T=120 \, MeV$ this expression
 gives the shift - 23  MeV. 
This is to be compared to -10 MeV in average low energy hadroproduction
reactions (\ref{eq_hadroproduced}) and to a shift of about -30 MeV
in  preliminary STAR  pp data \cite{STAR_Fachini} mention in the Introduction.

The important point here is that this effect is $quadratic$ 
in width. For example, the 3 times more narrow  $K^*$
is predicted to be shifted by this effect
an order of magnitude less.   
In contrast to that, much wider resonances -- e.g. 
the famous $\sigma$-meson, now listed   as $f_0(600)$ \cite{RPP}
with a width of about 300 MeV -- are changed beyond recognition.
 Applying the same expression we see that in the pp case 
a Boltzmann factor at  $T\sim 160 \, MeV$ transform it into a very
wide structure (which is indeed seen in hadroproduction), while at
kinetic freezeout
of heavy
ion conditions $T\sim 100 \, MeV$ it 
peaks at very low masses instead, see Fig,\ref{fig_Boltzmann}(b). 
(Since it is an s-wave, not p-wave resonance as
$\rho$, its  threshold suppression is weaker.)

We argued above that in the pp case the 2 pions forming the $\rho$ come from
a heat bath, while the outgoing pions propagate
essentially in free space. In contrast to that, 
in heavy ion collisions the local expansion rates are  slow
compared to the $\rho$ lifetime, and therefore it decays in matter.
So, in principle one
should  include
the final state factor  $(1\pm f)^2$. Direct experimental access to the magnitude of thermal
occupation rate at freezeout is provided by the integrated HBT correlator.
For $\rho$ it is however 
a small effect modifying the width by about 10 percent and
not affecting the mass. It is larger for $\omega$: see discussion
of it in \cite{Shu_pots}.

\begin{figure}[h!]
\centering
\includegraphics[width=6cm]{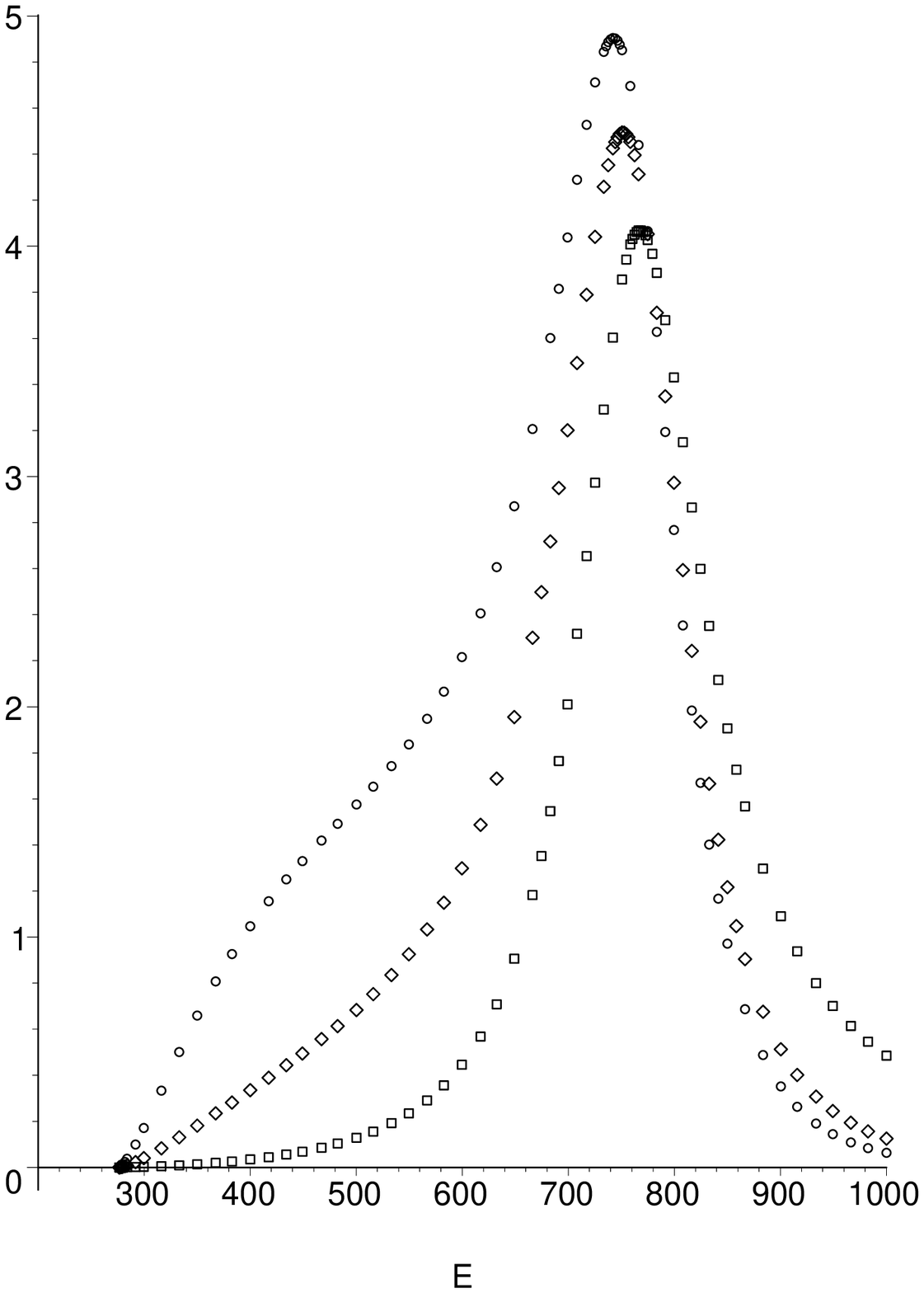}
\includegraphics[width=6cm]{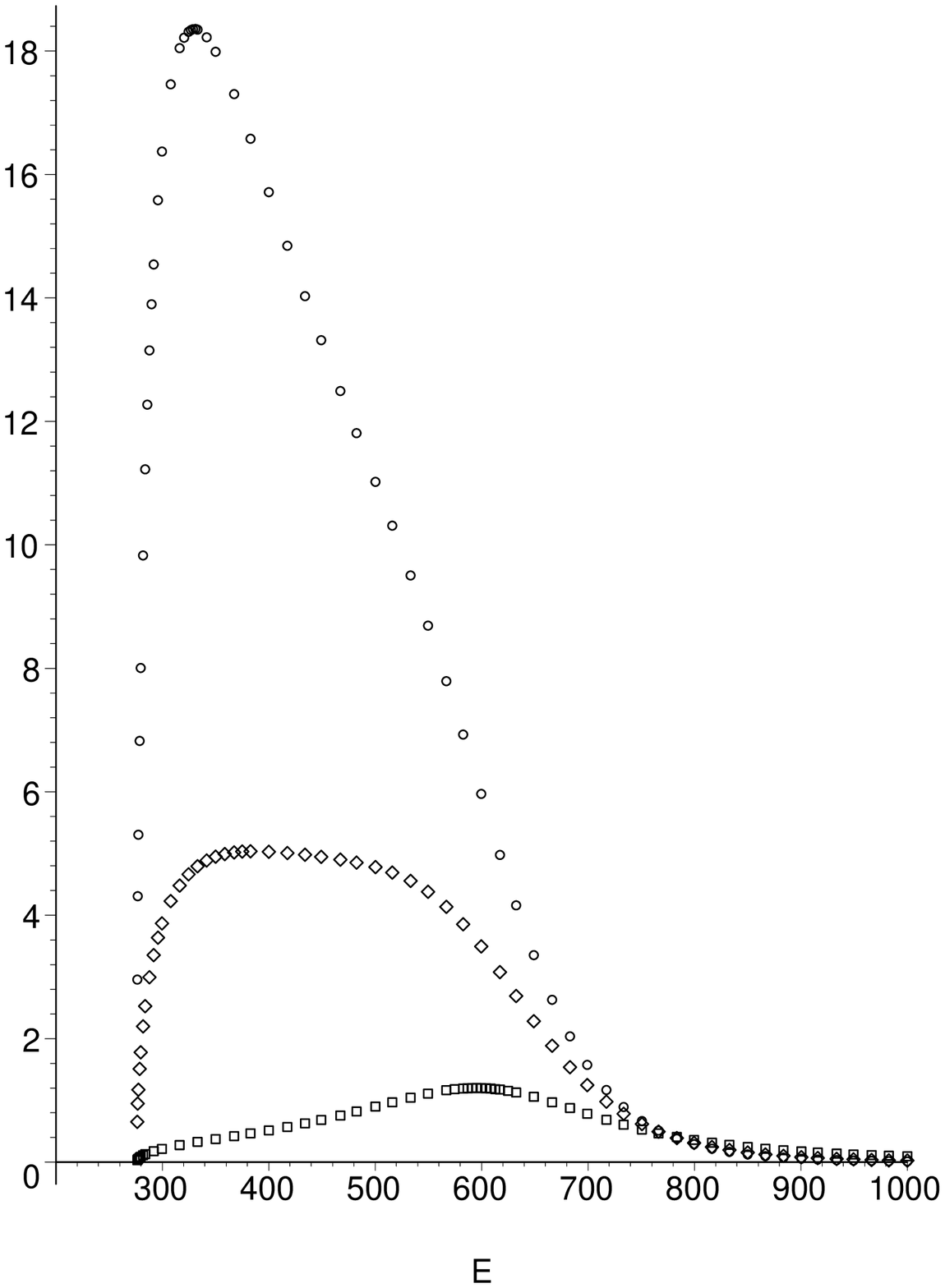}
\caption{\label{fig_Boltzmann} 
The shape of the $\rho$ (a) and $f_0(600)$
(b) as they appear  in elementary reactions
(squares), modified by a Boltzmann factor at chemical freezeout
T=165 MeV (diamonds)
and at chemical freezeout (circles)
}
\end{figure}

\subsection{Modification due to rescattering on pions, kaons and nucleons}

The formulae are all well known \cite{Shu_pots}, 
for the record we use the resonance
2-pion formfactors in the vector and scalar channels, which include
the phase space and $p^2$ in the p-wave matrix element in the width 
\be  \mathit{FF_{sq}} := {\displaystyle \frac {m^{4}\,}{(m^{2} - E^{
2} + f)^{2} + m^{2}\,\mathit{\Gamma_v}^{2}}}  \ee 
\be \mathit{FF_{sigma}} := {\displaystyle \frac {m^{4}}{(m^{2} - E^{2}
 + f)^{2} + m^{2}\,\mathit{\Gamma_\sigma}^{2}}} \ee
where
\be \mathit{\Gamma_v} := {\displaystyle \frac {\Gamma \,m\,k^{3}}{E\,
\mathit{kv}^{3}}}\ee 
\be \mathit{\Gamma_\sigma} := {\displaystyle \frac {\Gamma \,m\,k}{E\,
\mathit{kv}}} \ee

\be k := {\displaystyle \frac {1}{2}} \,\sqrt{E^{2} - 4\,\mathit{m_\pi}
^{2}} \ee
\be \mathit{kv} := {\displaystyle \frac {1}{2}} \,\sqrt{m^{2} - 4\,
\mathit{m_\pi}^{2}}\ee

We found that the resonance mass shifts due to s-channel resonances
depend  rather weakly
on the emission time, making predictions rather stable.
In particular, the difference between  
 (\ref{cond_res}) and  (\ref{cond_kinetic}) is at the level of 10-15
percent, much less than density difference which is  40 percent. This is
explained by a specific shape of the real part, with negative and
positive contributions weighted with a thermal weight. At lower
temperature higher states contribute less and their cancellation is 
less pronounced.

As seen from the  Table, the interaction 
of $\rho$ with the pions lead to  
negative shift of the order of -30 MeV, while with kaons it is basically zero.

Due to $N^*(1520)$ the interaction with nucleons leads to positive
 shift. There are however resonances above the threshold which
are not included; those will somewhat reduced our estimate. 
On the other hand, due to S=I=3/2 of the $\Delta$ state, it has rather large
statistical factor and significant
population at the relevant stage.  There
are sub-threshold $\rho-\Delta$ resonances in the 2 GeV mass region and thus
one can expect an effect similar in sign and 
magnitude to those from $N^*(1520)$, further increasing the upward
 shift.

An opportunity to test the role of $N\rho$ interacton 
is provided by heavu ion collisions
at lower  collision energies, of the order $\sqrt{s}$ few  GeV/N 
(like expected in new GSI proposed heavy ion project). Since
tha
increases the baryon fraction by an order of magnitude, 
the combined effect of $\rho-N,\rho-\Delta$ resonances can
compensate or even overcome the negative shifts discussed above.
The measurement of the exact mass and shift of $\rho$ in $\pi\pi$
mode at such conditions is therefore of significant interest.

\begin{table}[h!]
\caption{A set of resonances considered}
\centering
\begin{tabular}{lcrr}\hline
\it Name/Mass &\it Width &\it Branching & Mass Shift  \cr\hline
$a_1(1260)$ & 400 & 0.6 & -19 \cr
$a_2(1320)$ & 104 & 0.7 &  -15\cr
$K_1(1270)$ & 90 & 0.4 & +1.6 \cr
$K_2(1430)$ & 100 & 0.087 & -0.4 \cr
$N^*(1520)$ & 370 & 0.2 & 10 \cr
%$\Delta(1200)$ & & & 5 \cr 
\hline
\end{tabular}
\label{table1a}
\end{table}

\subsection{The effect of mutual attraction}
As explained above, the majority of the particles in the matter
are Goldstone bosons $\pi,K,\eta$ which do not
produce  attraction due  to
R-exchanges.
There should however be  an effect 
due to other particles, such as vector mesons and baryons. 
Unlike the s-channel resonances discussed above, the t-channel
attraction
scales as density, and thus is 40 percent stronger at
(\ref{cond_res}) as compared to  (\ref{cond_kinetic}).
Using a simple
expression for
the
 mass shift (\ref{eq_Rmassshift})  and the R-scalar mass $M_R\approx
 800 MeV$ one gets
\be \delta m_\rho^N \approx -28 MeV\ee
(in agreement with that  from the BR scaling)
due to all
 $\bar B+B$. An additional shift of the magnitude of
\be \delta m_\rho^v \approx -10 MeV\ee
comes from scalar R-exchanges between $\rho$ and 
 vector mesons $\rho,\omega,K^*$.

\subsection{Modification of the $rho$ width}
Finally let us briefly address the issue of the width modification.
The main difference between the two mechanisms of the mass shift
discussed above is that the t-channel attraction has no imaginary part and
is not associated with the broadening, while the
s-channel resonances lead to well defined
broadening, of the order of the sum of the absolute values of the mass
shifts
which is about 50 MeV. 

On the other hand, there is a ``kinematic'' effect working to the
opposite direction. The
negative mass shift discussed above automatically reduces the width,
both because of the reduced phase space and also due to the power of p
in the P-wave matrix element. The magnitude of this effect for
the predicted mass shift is
\be \delta \Gamma_\rho = 3{\delta m_\rho \over m_\rho}\Gamma_\rho \approx -50 \, MeV \ee

So, inside the accuracy one may claim at the moment, these two effects
seem to cancel each other.

\section{Discussion and Outlook}

Summarizing the paper, we have found that: (i) a  simple
initial state occupation effect 
is sufficient to explain the $\rho$ mass shift 
between pp and leptoproduction; (ii) the effect is 
predicted to be quadratic in width, so
it should be an order of magnitude smaller for $K^*$.
At the same time it leads to very significant shape deformation
for the sigma
$f_0(600)$ pushing it downward toward the threshold;
(iii) we found that all resonances are emitted at some time
between chemical and kinetic freezeout, at $T\approx 120 \, MeV$; (iv)
dynamical effects due to interaction of the $\rho$ with surrounding matter
 lead to about -50  MeV additional $\rho$
mass shift,
relative to a shift in  pp mentioned above.

All these conclusions about the mass shift of the $\rho$ agree with 
 the first STAR 
data shown at Quark Matter 02 \cite{STAR_Fachini}.
Hopefully quantitative results, with  fitted values of mass and width
versus centrality, will
be
soon available for more quantitative comparison.

The main uncertainties of our calculations of s-channel resonances
come from poorly known branching ratios for modes containing the $\rho$
mesons we focus on. It is clear that smaller contributions of
many more resonances are missing, which may or may not average out to zero.
We think the accuracy of our predictions are therefore at the level of
20-30 percent. Observation of similar phenomena with several
resonances
and at variable conditions (such as $N/\pi$ ratio) would help to
reduce this uncertainty. Very interesting would be precise
measurements
of the line shape of two more vector 
mesons, $K^*$ and $\phi$, and possibly a shape of $\omega$ in 3 pion channel.

In this paper we have not addressed in detail the issue of the $\rho$
$width$ and its broadening: however
as data would become more accurate, it would provide an
opportunity  to better test the relative role of s-channel and
t-channel conntributions to the mass shift.

STAR data also display a
 peak in $\pi\pi$ channel below $\rho$, at invariant mass $M\sim 600\, MeV$.
It can be partly $\sigma=f_0(600)$, which is however predicted in this
 work
to be deformed and move downward.
It is likely to be 2/3 of the $\omega$ with the $\pi^0$
missing, or a trace of baryonic
resonance like $N^*(1520)$.

\begin{acknowledgments}
We thank Patricia Fachini for her talk which started this work,
and several useful discussions. G.E.B. would like to thank Mannque Rho
for useful discussions. We also thank Ralf Rapp and Peter Kolb who shared
with us details of their unpublished works about hadronic matter at 
freezeout. 
This work was partially supported by the US-DOE grants DE-FG02-88ER40388
and DE-FG03-97ER4014.
\end{acknowledgments}

\end{narrowtext}

\begin{thebibliography}{99}
\bibitem{STAR_Fachini} P. Fachini (nucl-ex/0211001),
Talk at Quark Matter 2002, Nantes,
July 2002, to be published in proceedings in Nucl.Phys.A.  
\bibitem{STAR_Yamamoto} E.Yamamoto, $ibid$
\bibitem{Lolos} G.J.Lolos et al, Phys.Rev.Lett. 80 (1998) 241
\bibitem{Huber} G.M.Huber, G.J.Lobos and Z.Papandreou,
Phys.Rev.Lett. 80 (1998) 5285
%\bibitem{MEricson}
\bibitem{Shu_pots} 
E.~V.~Shuryak,
%``Collective Interaction Of Mesons In Hot Hadronic Matter,''
Nucl.\ Phys.\ A {\bf 533}, 761 (1991).
%%CITATION = NUPHA,A533,761;%%
\bibitem{Shu_Tho} E.~V.~Shuryak and V.~Thorsson,
%``Kaon modification in hot hadronic matter,''
Nucl.\ Phys.\ A {\bf 536}, 739 (1992).
%%CITATION = NUPHA,A536,739;%%
\bibitem{Kap_etal}
V.L. Eletsky, M. Belkacem, P.J. Ellis,
J.I. Kapusta,Phys.Rev.C64:035202,2001 ;
 nucl-th/0104029
\bibitem{RG}R.~Rapp and C.~Gale,
%``rho properties in a hot meson gas,''
Phys.\ Rev.\ C {\bf 60}, 024903 (1999)
[arXiv:hep-ph/9902268].
%%CITATION = HEP-PH 9902268;%%
\bibitem{CRW}R.~Rapp and J.~Wambach,
%``Chiral symmetry restoration and dileptons in relativistic heavy-ion  collisions,''
Adv.\ Nucl.\ Phys.\  {\bf 25}, 1 (2000)
[arXiv:hep-ph/9909229].
%%CITATION = HEP-PH 9909229;%%
\bibitem{BR_scaling} G.~E.~Brown and M.~Rho,
``Matching the QCD and hadron sectors and medium dependent meson masses:  Hadronization in relativistic heavy ion collisions,''
arXiv:nucl-th/0206021 and references therein.
%%CITATION = NUCL-TH 0206021;%%
\bibitem{CEG}
G.~Chanfray, M.~Ericson and P.~A.~Guichon,
%``Chiral symmetry and quantum hadro-dynamics,''
Phys.\ Rev.\ C {\bf 63}, 055202 (2001)
[arXiv:nucl-th/0012013].
%%CITATION = NUCL-TH 0012013;%%


\bibitem{BR} G.E.Brown and M.Rho, Phys.Rept.269 (1996) 334;
Nucl.Phys.A596 (1996) 503.
\bibitem{Shu_80}E.~V.~Shuryak,
%``Quantum Chromodynamics And The Theory Of Superdense Matter,''
Phys.\ Rept.\  {\bf 61}, 71 (1980), Appendix A.
%%CITATION = PRPLC,61,71;%%
\bibitem{Leutwyler}J.~Gasser and H.~Leutwyler,
%``Thermodynamics Of Chiral Symmetry,''
Phys.\ Lett.\ B {\bf 188}, 477 (1987).
%%CITATION = PHLTA,B188,477;%%
P.~Gerber and H.~Leutwyler,
%``Hadrons Below The Chiral Phase Transition,''
Nucl.\ Phys.\ B {\bf 321}, 387 (1989).
%%CITATION = NUPHA,B321,387;%%
\bibitem{DEI} M.Dey, V.L.Eletsky and B.L.Ioffe, Phys.Lett. B252 (1990) 620,
\bibitem{EI}V.~L.~Eletsky and B.~L.~Ioffe,
%``Next-to-leading order temperature corrections to correlators in QCD,''
Phys.\ Rev.\ D {\bf 51}, 2371 (1995)
[arXiv:hep-ph/9405371].
%%CITATION = HEP-PH 9405371;%%
\bibitem{Pisarski} R.~D.~Pisarski,
%``To VMD, or not to VMD, in the quark - gluon plasma,''
arXiv:hep-ph/9505257,9503328,9503329.
%%CITATION = HEP-PH 9505257;%%
\bibitem{DL}E.~G.~Drukarev and E.~M.~Levin,
%``The QCD Sum Rules And Nuclear Matter. 2,''
Nucl.\ Phys.\ A {\bf 511}, 679 (1990)
[Erratum-ibid.\ A {\bf 516}, 715 (1990)].
%%CITATION = NUPHA,A511,679;%%
\bibitem{RPP} Review of Particle Properties, Phys.Rev.D66  (2002) 010001
\bibitem{Beccatini}
F.~Becattini,
%``Statistical hadronization phenomenology,''
Nucl.\ Phys.\ A {\bf 702}, 336 (2002)
[arXiv:hep-ph/0206203].
%%CITATION = HEP-PH 0206203;%%
\bibitem{HS}
C.~M.~Hung and E.~V.~Shuryak,
%``Equation of state, radial flow and freeze-out in high energy heavy ion  collisions,''
Phys.\ Rev.\ C {\bf 57}, 1891 (1998)
[arXiv:hep-ph/9709264].
%%CITATION = HEP-PH 9709264;%%
\bibitem{Bebie}
H.~Bebie, P.~Gerber, J.~L.~Goity and H.~Leutwyler,
%``The Role of the entropy in an expanding hadronic gas,''
Nucl.\ Phys.\ B {\bf 378}, 95 (1992).
%%CITATION = NUPHA,B378,95;%%
\bibitem{Teaney}D.~Teaney,
%``Chemical freezeout in heavy ion collisions,''
arXiv:nucl-th/0204023.
%%CITATION = NUCL-TH 0204023;%%
\bibitem{Rapp}R.~Rapp,
%``Hadro-chemistry and evolution of (anti-) baryon densities at RHIC,''
Phys.\ Rev.\ C {\bf 66}, 017901 (2002)
[arXiv:hep-ph/0204131].
%%CITATION = HEP-PH 0204131;%%
\end{thebibliography}
\end{document}